\documentclass{llncs}
\usepackage{makeidx}  % allows for indexgeneration
\usepackage{multirow}
\usepackage{graphicx}
\begin{document}
\frontmatter          % for the preliminaries
\pagestyle{headings}  % switches on printing of running heads
\addtocmark{Hamiltonian Mechanics} % additional mark in the TOC
\mainmatter              % start of the contributions
\title{A Factorial Experiment on Scalability of Search Based Software Testing}
%
%\titlerunning{Hamiltonian Mechanics}  % abbreviated title (for running head)
%                                     also used for the TOC unless
%                                     \toctitle is used
%
\author{Arash Mehrmand , Robert Feldt}
%
%\authorrunning{Ivar Ekeland et al.} % abbreviated author list (for running head)
%
%%%% list of authors for the TOC (use if author list has to be modified)
%\tocauthor{Ivar Ekeland, Roger Temam, Jeffrey Dean, David Grove,

%
\institute{ Blekinge Institute of Technology, SE-371 79 Karlskrona, Sweden}
%\email{arme08@student.bth.se}
%\email{robert.feldt@bth.se}
%\texttt{http://users/\homedir iekeland/web/welcome.html}

\maketitle              % typeset the title of the contribution

\begin{abstract}
Software testing is an expensive process, which is vital in the industry. Construction of the test-data in software testing 
requires the major cost and to decide which method to use in order to generate the test data is important. 
This paper discusses the efficiency of search-based algorithms (preferably genetic algorithm) versus random testing, in software test-data generation. 
%A factorial experiment is designed so that, we have more than one factor for each experiment we make. 
%Although many studies have been done in the area of automated software testing,%
This study differs from all previous studies due to sample programs ot software under test (SUT) which are used. 
Since we want to increase the complexity of SUTs gradually, and the program generation is automatic as well, Grammatical Evolution is used to guide the program generation. 
SUTs are generated according to the grammar we provide, with different levels of complexity. 
SUTs will first undergo genetic algorithm and then random testing.
Based on the test results, this paper recommends one method to use for automation of software testing. 
%SUTs are not like the sample programs, provided by other studies since they are generated using a grammar.
\end{abstract}
\section{Introduction}
Software testing is one of the most critical and at the same time most time and cost consuming activity in software development process \cite{1}. In order to cut this costs as much as possible we need to employ effective techniques to automate this process. What actually needs to be automated is generating the test data which is a demanding process \cite{2}. Automation of generating test data helps to have fast, cheap and error free process \cite{2}.

If we can optimize the test cases generation process, then we save a lot of time and money. Among optimization techniques, search based optimization is the one which has been applied to different areas of Software Engineering applications \cite{2}.

Software testing was not an exception in Software Engineering area and the term ``Search Based Software Testing (SBST)`` proves it. The recent years show a dramatic rise of interest in Search Based Testing (SBT) \cite{2}. SBT has been applied to different testing areas such as structural, functional, non-functional, mutation, regression and so on \cite{2}.

The term "Search Based Testing" in general means using some sort of search alogorithms to generate test cases for testing a software. 

There are many Search Based techniques and each of them uses a different algorithm to optimize the search and the results.

The three popular ones which are in use for testing purpose are Neighborhood Search, Simulated annealing and Genetic algorithms (GA). And of course there are simple techniques as well such as random testing which does not use any optimization and search technique. \cite{3}. According to \cite{3} flexibility and strength of these heuristic optimization techniques help a lot in finding the optimal solutions in large and complex search spaces.

This study focuses on using the SBT techniques or as we said heuristic optimization techniques for software testing, when we have different programs regarding to complexity. We will apply the technique which is called Genetic algorithm to wide variety of programs to see the effectiveness of Genetic algorithm compared to random testing.
The final results of this paper could help Software engineers and especially Software testers to compare the efficiency of search based techniques versus random testing when it comes to get the maximum percentage of coverage on each code, as they deal with programs with different levels of complexity.

%As the final result this study will show that GA in software testing is  more effective than random testing as we increase the complexity of software.

The structure of this paper is as follows: Next section, presents the existing research in this area.
Section 3 explains the experimental approach of this study and study design. Section 4 talks about the design of experiment in this study. Sections 5 and 6 include the results and discussion about them. And finally conclusion and further work will come under section 7.
\section{Previous Research and Related Work}
There exist many studies in software testing automation area and SBST. Some of them deal with software complexity as well but we could have not found any of them similar to this study as we deal with increasing of software complexity along our study.
For example, authors in \cite{4} discuss the experiments with test case generation with large and complex programs and they conclude that, there is a wide gap between the techniques based on GA and those based on random testing. They have chosen some programs which are written in C language and used them as SUT.

Compared to what the authors tried to explain in \cite{4}, our study has a major difference and that is the use of many Java programs that vary in size and complexity.

Many other studies tried to experiment software testing using genetic algorithms and compare the results with other techniques. Authors in \cite{5} performed experiments on some small and a few complex test objects and they came up with this results: Particle Swarm Optimization overcomes GA in coverage of many code elements so not always GA is the best solution. Another paper also talks about GA based software testing and the main idea is to find problematic situations while testing programs with GA \cite{6}.

Alba and Chicano in their paper \cite{7} describe how canonical genetic algorithms and evolutionary strategies can help in software testing. They used a benchmark of twelve test programs and applied both Evolutionary Strategies and GA and compared the results at the end.

However, many existing research and studies (e.g. \cite{8},\cite{9},\cite{10}) have used SBT specially GA in order to test the software and compare the results with other techniques.
\section{Experiment Approach}
The chosen approach for this study is factorial experiment which compares \textit{GA} and \textit{random testing} in efficiency. 
Factorial experiment is an experiment whose design consists of two or more factors and each factor has discrete possible values which we call them levels \cite{1}.
Factors, mentioned in table 1, are the test search techniques which has GA and random as values, complexity which has low, medium and high as levels , statement coverage and branch coverage with their defined levels. Since we have to increase the complexity of our generated programs gradually, there are some factors which we have to increase or decrease, in order to change the complexity the programs. 
Code complexity is actually an independent variable in our experiment. Statement and branch are independent of each other but dependent to the complexity level; the more
 complexity level we have, the more branch or statement we deal with.
Below is a table which shows the factorial design of this experiment.
\begin{table}
\centering
\caption{Factorial Design of The Experiment}
\begin{tabular}{|l|l|l|l|l|} \hline
Factor's name&TST&Complexity&Statement&Branch\\ \hline
\multirow{3}{*}{Level}&GA&Low & 75& 25\\ 
&Random&Medium &150& 50\\
&&High &300& 100\\ \hline

\end{tabular}
\end{table}

For both statement and branch coverage, we have source codes with low, medium and high complexity. For each coverage method, we test our codes with both GA and random testing and record the coverage percentages as you can see in plots under section 5.

These factors (variables) are explained in details in section 4 which is Experiment Design. After making all the experiments, we will have two tables and six graphs which show the efficiency of GA compared to random testing; these results will come in section 6 which is Experiment Results.
\section{Experiment Design}
In order to design a fully automated process for our experiment, we need to follow a step by step structure. These steps are the following subsections:
%\begin{enumerate}
 %\item Generate the SUTs with different size and complexity automatically using grammatical evolution \cite{12}.
 %\item Apply GA to the generated SUTs in order to test them.
 %\item Instrument, compile, and run the written unit test cases.
 %\item Apply random testing to SUTs.
 %\item Getting the coverage report which shows the percentage in which we have covered our code using both GA and random testing so we can compare them.
%\end{enumerate}
%Now all mentioned items above, will be described in order to clarify the experiment design.
\subsection{Generate Software Under Test (SUT) Using Grammatical Evolution (GE)}
Grammatical Evolution (GE) is a kind of Genetic Programming based on grammar. GE combines basis and rules of molecular biology with representational power of formal grammars. GE has a unique flexibility due to its rich modularity, and makes it possible to use alternative search strategies, whether evolutionary, or other heuristic,    be it stochastic or deterministic, and to radically change its behavior by only slight changes to the grammar supplied. One of the main advantages of GE, is the easiness of modifying the output structures by simply editing the plain text grammar. This advantage actually originates from the use of grammar to describe the structures that are generated by GE \cite{12}.

To use GE, for our SUTs generation, we used an open source software implementation of grammatical evolution in Java. Grammatical Evolution in Java (GEVA) is the name of the tool which provides the possibility to use GE with our supplied grammar, in order to generate evolved programs. However, for us, to use GEVA, there was a need to make some modifications and supply our own material to reach our goal which is generating Java programs with different levels of complexity.

First of all we need to define our problem, which is generating different Java programs. For this purpose, we wrote a Java code which automatically generates the header and footer of the programs and makes the contents of the body of the program, according to the BNF which we will provide. So next step is to write a BNF. Using Backus-Naur form (BNF), it is possible to provide the specification of a programming language structure, which is Java in this case. Since we have to have a subset of grammars, we should define our BNF according to what subsets of Java programming language we need to generate. The genetic configuration of GEVA is also changed according to our need. Following is a part of our properties, used in GEVA:
\begin{itemize}
 \item Population size: 200 
 \item Generation: 10000
 \item Initial chromosome-size: 200
 \item Selection type: Roulette wheel \cite{17}.
\end{itemize}

Providing GEVA, with the problem definition and BNF file, it can generate our desired programs. Because we want to increase the complexity of our programs, we need to run the generation, three different times. When the fitness function is set to the number of statements, we run the generation, once with 75 statements, once with 150 and finally with 300 statements. And when the fitness function is set to the number of branches in the code, we run the generation once with 25 branches, once with 50 branches and finally with 100 branches. However, the complexity levels could be extended to more than three levels, e.g. very low, low, medium, high, very high or we can have the same three levels with different numbers; the results and the effectiveness of GA over random search will not change though. So three levels which are used in this experiment are sufficient enough for us to conclude which method is more efficient.

It is worth mentioning that our sample programs are just regular codes written in Java and they involve different kinds of conditions and statements.
By this we mean we did not follow any specific problem to generate the sample programs.
\subsection{GA}
After generating the SUTs, we need to apply GA to our code so it will act genetically during the testing process. The library which is used for this purpose, in this study, is called JGAP. JGAP is actually a Genetic Algorithm and Genetic Programming component which is provided as Java framework. Using JGAP it is possible to apply evolutionary principles to problem solutions since it provides us, the basic genetic mechanisms \cite{13}.

Using JGAP, first step is to plan the chromosomes, so we have to decide on type and quantity of genes that we are going to use. Each chromosome is made of genes, and in this study each gene is actually a test data. And depending on the size and number of input variables, we decide on number of genes. Second step is to implement fitness function. Although JGAP is designed to do almost all the evolutionary steps in different problems but it does not know, by itself, whether or not a potential solution is better than the other ones. Our fitness function, depending on which part of experiment we are, is the coverage percentage of the generated test data. For example if our complexity metric is the number of statements, then the best chromosomes are those which provide better statement coverage on the code. Third step is to setup the configuration object of JGAP which means deciding on how big our population is, and all other GA configurations. In this experiment the \textit{population size} is set as 200 and \textit{number of generations} is set to 10000. Fourth and fifth steps are creating and evolving the population. And obviously after each generation, the best test data of each population are taken. However, the followings are the steps, GA basically, uses to test a software where $P(t)$ is the population and $t$ is the generation \cite{17}:
\begin{verbatim}
 Initialize P(t)
  Evaluate P(t)
  While the termination condition
 	is not reached, do
		select P(t+1) from P(t)
		recombine P(t+1)
		evaluate P(t+1)
		t=t+1		     
\end{verbatim}              
Evaluating the fitness function depends on which experiment we are running; so for statement coverage and branch coverage it differs. Please refer to 3.1.3, where defining the fitness of the experiment is explained in more details.
This was the whole strategy of GA application to our SUTs.
\subsubsection{Defining The Fitness Function}
What genetic algorithm does is actually searching for the suitable test case, to kill a given mutant; for this purpose fitness function is used, which assigns a non-negative cost to each candidate input value. The more cost each input value has, the more appropriate it is, so we aim to maximize this value and therefore maximize the fitness function \cite{21}.

%According to \cite{21} there are three kinds of conditions which can be used as fitness function; in the other words, there are three kinds of costs which we can be maximized to maximize the fitness function. However, the combination of these costs can be used as well. Those conditions are called: reachability condition, necessity condition, and sufficiency condition. Reachability condition is actually the same issue as function minimization. In this case a numerical predicate is assigned to each branch. If the value of the branch predicate is equal to zero, this means that the condition has been satisfied and the fitness function is maximized. But in a case that two candidate inputs have a non-zero predicate value, then to compare them we need to pick the one which has the closer value to satisfy the common failed branch predicate.

%The other two conditions are not used in our fitness function , this is why they are not defined.

We used the term which is called \textit{function minimization} \cite{17}. Function minimization helps to find the desired input for each condition to be executed. Using this way, each condition in the code has its own function. Genetic search, then, tries to find the input which minimizes the value of that specific function. For example, imagine we have this condition on line 300 of the program:
\begin{verbatim}
 if (x >= z+10)
\end{verbatim}
and we aim to execute the true condition of this branch. Then the, the minimization function is defined as below:

 \[f(n) = \left\{ 
\begin{array}{l l}
 (z_{300}+10)-x_{300} & \quad if Z_{300}+10 \leq  x_{300}\\
  0 & \quad \mbox{otherwise}\\
\end{array} \right. \]

The above function says, if $z_{300}+10$ is less than or equal to $x_{300}$ then try to set the value of $z_{300}$ and $x_{300}$ so that the value of $(z_{300}+10) - x_{300}$ is as minimum as possible. Please notice that $300$ here is the line number. We have to define the line number since the value of input parameters (x and z), change over time, as the programs run.

When the fitness function is statement coverage, we need to count the number of statements which are blocked between braces of correspondent branch. We use it as the weight and divide the value of [the correspondent] function with this weight. For branch coverage, we do not need this weight since we are looking for the covered branches not the statements.

Because our sample programs execute repeatedly over time, the input parameters change every time the program executes. The minimization function decides, how closed the input parameters are to the desired values which satisfy the conditions and we use GA in order to minimize the function; the fittest chromosomes mean the closer results to minimize the function.

\subsection{Random Testing}
Random testing is the chosen approach in this study, to be compared with SBST. According to \cite{16} random testing is actually the use of randomly generated of test data which has some advantages such as easy and simple implementation and speed of execution, which means less run time. When running the test using random testing, we have to define a parameter; this parameter defines how many times we should generate test data using random testing. For example if we set the parameter to 100, then it generates 100 set of test cases. Out of these [e.g.] 100, we take the best and compare it with the fittest result of GA. However, in this experiment, this number is set to $100000$.

Using random testing we need to define our input domain, for test data, so the numbers will be randomly generated form this domain. In this study this set starts from $- 1000000 $ to $+ 1000000$.

\section{Experiment Results}
After testing the generated SUTs, we came up with the results. As you can see in the following plots, GA outperforms the random testing in almost, all the experiments which we have done. As mentioned before, the whole experiment includes testing 60 programs; 30 (10 with low, 10 with medium, and 10 with high complexity) with statement coverage purpose and 30 (10 with low, 10 with medium, and 10 with high complexity) with branch coverage purpose. So the fitness function was different for each purpose. The generated programs are also different since they are generated with two different fitness function. The sample programs for statement coverage purpose, are generated with number of statements purpose and the ones which are used for branch coverage purpose, are generated with number of branches purpose. Please not that the following plots include 10 programs each, which are randomly chosen from multiple runs that we had.

\begin{figure}[ht]
\begin{minipage}[b]{0.5\linewidth}
\centering
\includegraphics[scale=0.4]{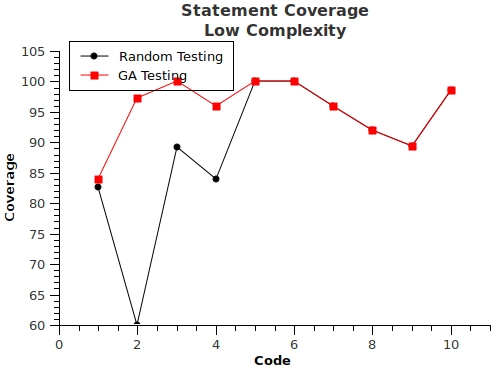}
\caption{GA vs Random-low complexity}
\label{fig:figure1}
\end{minipage}
\hspace{0.5cm}
\begin{minipage}[b]{0.5\linewidth}
\centering
\includegraphics[scale=0.4]{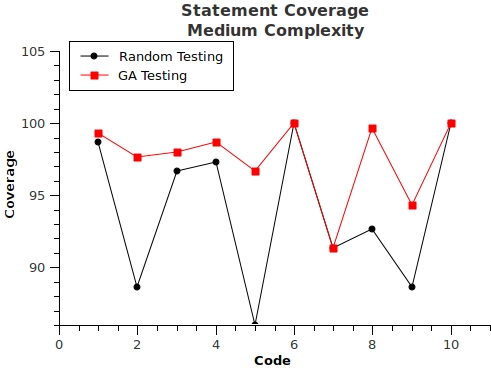}
\caption{GA vs Random-medium complexity}
\label{fig:figure2}
\end{minipage}
\end{figure}
\begin{figure}[ht]
\begin{minipage}[b]{0.5\linewidth}
\centering
\includegraphics[scale=0.4]{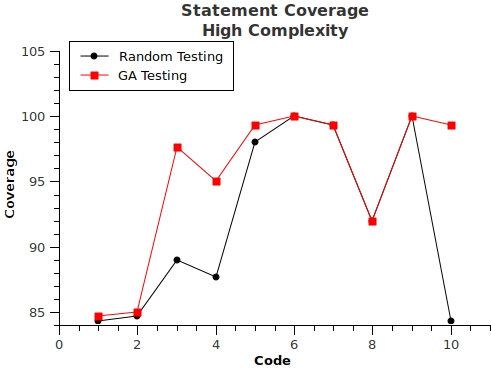}
\caption{GA vs Random-high complexity}
\label{fig:figure1}
\end{minipage}
\hspace{0.5cm}
\begin{minipage}[b]{0.5\linewidth}
\centering
\includegraphics[scale=0.4]{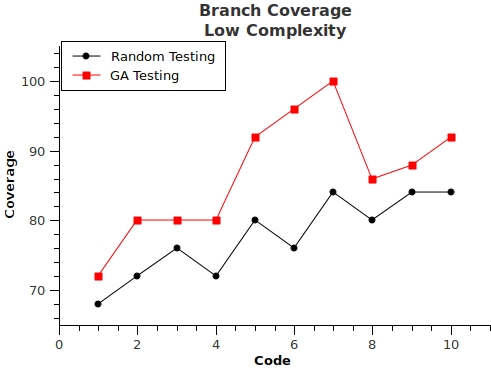}
\caption{GA vs Random-low complexity}
\label{fig:figure2}
\end{minipage}
\end{figure}
\begin{figure}[ht]
\begin{minipage}[b]{0.5\linewidth}
\centering
\includegraphics[scale=0.4]{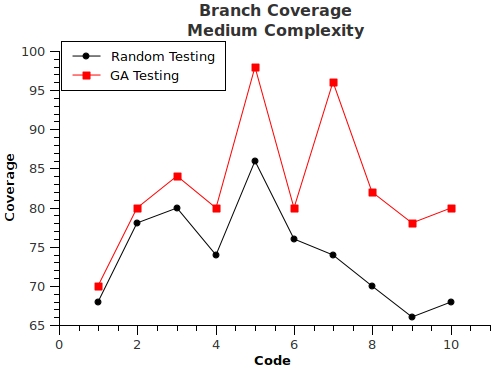}
\caption{GA vs Random-medium complexity}
\label{fig:figure1}
\end{minipage}
\hspace{0.5cm}
\begin{minipage}[b]{0.5\linewidth}
\centering
\includegraphics[scale=0.4]{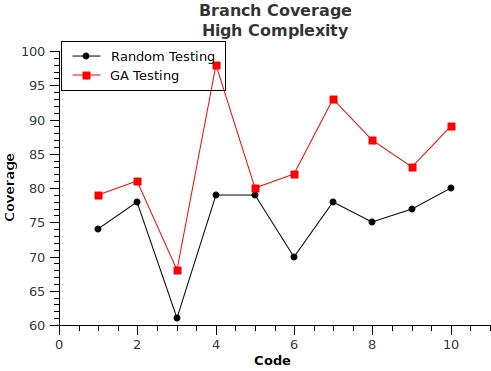}
\caption{GA vs Random-high complexity}
\label{fig:figure2}
\end{minipage}
\end{figure}
%\begin{figure}
%\centering
%\includegraphics[scale=0.45]{plot.jpeg}
%\caption{GA vs. Random for low complexity}
%\includegraphics[scale=0.45]{plot_1.jpeg}
%\caption{GA vs. Random for medium complexity}

%\end{figure}
%\begin{figure}
%\centering
%\includegraphics[scale=0.6]{plot_1.jpeg}
%\caption{GA vs. Random for medium complexity}
%\end{figure}
%\begin{figure}
%\centering
%\includegraphics[scale=0.6]{plot_2.jpeg}
%\caption{GA vs. Random for high complexity}
%\end{figure}
%\begin{figure}
%\centering
%\includegraphics[scale=0.6]{plot_3.jpeg}
%\caption{GA vs. Random for low complexity}
%\end{figure}
%\begin{figure}
%\centering
%\includegraphics[scale=0.6]{plot_4.jpeg}
%\caption{GA vs. Random for medium complexity}
%\end{figure}
%\begin{figure}
%\centering
%\includegraphics[scale=0.6]{plot_5.jpeg}
%\caption{GA vs. Random for high complexity}
%\end{figure}

As mentioned above, these plots represent the result of 10 chosen programs, but to see how efficient GA is, please refer to the tables II and III, where they show the average percentage of coverage with both GA and random search.

\begin{table}
\centering
\caption{Factors and their levels}
\begin{minipage}{0.5\textwidth}
\begin{tabular}{|c|c|c|c|c|c|l|} \hline
&&\multicolumn{4}{|c|}{Test Search Technique}&\\ \hline
 & & \multicolumn{2}{|c|}{GA}& \multicolumn{2}{|c|}{Random} &\\ \hline
Cmplx\footnotemark[1] &Stmt\footnotemark[2]& Mean&Stdev&Mean&Stdev&Actual CL\footnotemark[3]\\ \hline
Low &75& 95.87&12.13&93.87&13.95&44.42\\ \hline
Medium &150& 96.12&8.74&94.73&8.98&45.41\\ \hline
High &300& 95.39&5.63&94.13&6.16&58.84\\ \hline
\end{tabular}
\footnotetext[1]{complexity}
 \footnotetext[2]{statements}
 \footnotetext[3]{Confidence Level}
\end{minipage}
\end{table}
In table II, for statement coverage, from low to high, t-test shows that means are not significantly different from the Selected Confidence Level. Although we might not be able to prove GA outperforms random search statistically because of the actual confidence level here, but still the overall mean of coverage in GA is more than random search in all cases.

Note: desired confidence level is set to 90

\begin{table}
\centering
\caption{Factors and their levels}
\begin{minipage}{0.5\textwidth}
\begin{tabular}{|c|c|c|c|c|c|l|} \hline
&&\multicolumn{4}{|c|}{Test Search Technique}&\\ \hline
 & & \multicolumn{2}{|c|}{GA}& \multicolumn{2}{|c|}{Random} &\\ \hline
Cmplx\footnotemark[4] &Branch& Mean&Stdev&Mean&Stdev&Actual CL\footnotemark[5]\\ \hline
Low &25& 87.27&6.98&82.4&7.62&98.76\\ \hline
Medium &50& 79.33&6.99&75&5.82&98.84\\ \hline
High &100&76.57&8.25&73&5.76&94.32\\ \hline
\end{tabular}
\footnotetext[4]{complexity}
 \footnotetext[5]{Confidence Level}
\end{minipage}
\end{table}

And in Table III, for branch coverage, from low to high, t-test shows that means are significantly different from the Selected Confidence Level.

Note: desired confidence level is set to 90
\section{Discussion}
the results from the plots in section 6, illustrate GA has outperformed random testing in most cases and in very few cases they have same coverage percentage.

The design of this experiment is factorial so that we can assure in different situations, with different complexity levels and different source codes, random testing can never outperform GA from the coverage and efficiency point of view.

However, there exist some issues which need to be discussed here, some of them could be serious problems though, if they are not handled properly.

Our SUTs here are generated using GE. As mentioned in previous sections, for GE to generate programs, we have to provide it with a BNF. Unfortunately it is impossible to define semantics (logic) in BNF file which means there is a very low chance of controlling what you are generating. 
During these experiments, we had many programs which had same statements, or same loops or even same blocks due to GA behavior of generating fittest chromosome. 
%Since we follow GA in our grammatical evolution, whenever it finds a fit chromosome, it keeps using it to generate the new population since it is the fittest. What we mean by chromosome here is actually the statements inside the code. For example if the purpose is number of statements, then if GEVA finds a block a statement which satisfies the fitness function, it continues to use it more and more. So at the end we will have a code which contains some repetition.%

Big numbers, which are made throughout the process of multiplication, addition and subtraction, are out of range of integer and double variable. These out of range variables can cause the code not to be tested completely because when the test reaches these numbers, stops running. We did not define divisions in our BNF since it can cause many problems e.g. division by zero.

Another important issue is the infinite loops in the code. And , we controlled them using exit-loop which means having maximum iteration and break it after the value of maximum iterations have been reached. However, having breaks, even if we assign maximum iteration to a large number, causes the code not to get tested completely \cite{14,15}.

Conditions in the loops are very important during testing, because if we can satisfy the condition and go through the block of that condition, we can execute more statements. There are two kind of conditions which can cause problems. First one is the condition which is not reachable at all and second kind is a very easy condition which both GA and random testing can cover it easily.

%As an example, have a look at the following code, when GA can never cover it; and of course when GA cannot cover it, random testing cannot do it either.

%\begin{small}\begin{verbatim}
% while ( ( z + 6 7 2 ) <= ( z - 6 7 5 4 ))
%
%AND

 %while ( ( y <= ( y + 5 3 7 ) )
%&& ( z >= ( z - 6 9 ) ) )
%\end{verbatim}
%\end{small}

This problem is solved in our BNF which means we have the minimum number of easy and unreachable conditions so we do not face serious problems during our experiment.

%Have a look at the source code below, which is a small piece from a source code.
%\begin{small}

%\begin{verbatim}

% if(((x+41168)<(z+4))&&((y+641)!=(x-3619))
%&&((x+36321)==(x+1691)))
%{
%x=(y-((z+x)-x));
%}
%if((z+25)<=(y+96131))
%{
%y++;
%x++;
%}
%else
%{
%x++;
%}
%\end{verbatim}
%\end{small}
%Let's say we want to test this code with the purpose of statement coverage. First loop will never run, no matter if GA is in use or random testing because the last condition of this loop makes it false all the time.
%$(x+36321)==(x+1691)$ will never become true and it make the whole loop not to run. So the statement block of this loop will never be reached. On the other hand, next loop which is $if((z+25)<=(y+96131))$ is so easy that both GA and random testing can cover it and of course the statement block of this loop will be reached with either GA or random testing. Now imagine a source code with hundreds of lines and filled with many of mentioned loops and blocks. In these cases, it does not really matter which method we are using to test our programs since we get the same coverage.

Using GE and BNF to generate programs has some difficulties and limitations which make it very hard to generate a piece of code without infinite loops, hard but possible to cover conditions. 
And testing the programs which are generated using GE has its own limitations since we have limited the number of iterations of existing loops, 
and eliminate the nested loops because of exit-loop problem. 
The authors in \cite{18} do not recommend using nested loops; instead they suggest another way which is called LoopN. 
In LoopN the use of nested loops is forbidden. They think loop does not do much harm if it is not nested. 
However limiting the code, and not having the nested loops can decrease the overall complexity, which was not our aim but there was no other way than eliminating them. 

Rather than having the problem with BNF, instrumentation itself, is a big issue. This is mainly because, the more details we need to instrument, the less speed we have. For example in case of nested loops; instrumenting nested loops needs us to keep track of each loop (inner and outer), each branch, and each condition more precisely.

\section{Conclusion}
In this paper, we have reported on results from a factorial experiment, where the effectiveness of GA is compared to random testing in automation of software testing. In this experiment, the process of generating SUTs were also automated, which means we used GE to generate our sample Java programs. To our knowledge, the presented results we had are not yet reported in the software test data generation area, because of having automatically generated programs, different levels of complexity and different kinds of coverage at the same time. 
However, it is worth mentioning, this study focuses on the efficiency of the techniques and does not consider the other attributes such as time, cost etc.
The followings are our observations from the experiments and the taken results.
\begin{itemize}
% \item Generating programs using GE, can limit the source code because of mentioned problems about BNF, loops, conditions etc. Of course GE is effective when we are looking for a solution of a problem. The studies which show the flexibility of GE in problem solving include: grammars have been used to represent a diverse array of structures including binary strings, code in various programming languages (e.g., C, Scheme, Slang, Postscript), music, financial trading rules, 3D surfaces, and even grammars themselves \cite{12}.
\item GA can outperform random testing and it did actually in this study with our automatically generated SUTs. We proved it using a factorial experiment. This factorial experiment with different levels of code complexity and two methods of code coverage, shows that in most of the situations GA has better efficiency than random testing.
\item Random testing could be, however, a faster solution but it is not recommended since in real life we look for a software with less bugs, not a software which is tested quickly. This means more bugs that we find in our programs, more cost we save; this is what happens in real world.
 \item Number of statements, is not a good metric for software complexity because adding extra statement hardly makes the code more complex. What makes the code really complex, is the conditions.

\end{itemize}

This study can be extended by applying well known software complexities i.e. cyclomatic complexity.

%Another area which is possible to work on, is to use GE with Extended BNF (EBNF). Using EBNF, it is possible to define not only the syntax, but also some additional control elements such as sequence, choice, option, and repetition. It does not solve the semantics problem though.

%If it is possible to develop a fast coverage tool (open source preferably) which includes the instrumentation as well, then many problems are solved in this area. They could be even more useful if they are open source since being open source can speed up the testing and getting reports by grouping the required classes as one unit package and compile the desire tool with only desired features.

%
% ---- Bibliography ----
%

%
% second contribution with nearly identical text,
% slightly changed contribution head (all entries
% appear as defaults), and modified bibliography
%

\begin{thebibliography}{5}
%
\bibitem {1}
Catelani, M.,, Ciani, L.,, Scarano, V. ,, Bacioccola, A.:
A Novel Approach To Automated 
Testing To Increase Software Reliability.
Instrumentation and Measurement Technology Conference Proceedings 1499--1502 (2008)

\bibitem {2}
Harman, Mark:
Automated Test Data Generation 
using Search Based Software Engineering.
AST '07: Proceedings of the Second International Workshop on Automation of Software Test, 2 (2007)

\bibitem {3}
Nigel James Tracy:
A Search-Based Automated Test-Data Generation
 Framework for Safety-Critical Software.
University of York (2000)

\bibitem {4}
Christoph C. Michael and Gary E. Mcgraw and Michael A and Curtis C. Walton:
Genetic algorithms for 
dynamic test data generation.
In Proc. ASE’97, 307--308 (1997)

\bibitem {5}
Windisch,, Andreas and Wappler,, Stefan and Wegener,, Joachim:
Applying particle swarm 
soptimization to software testing.
GECCO '07: Proceedings of the 9th annual conference on Genetic and evolutionary computation, 1121--1128 (2007)

\bibitem {6}
Alander, Jarmo T. and Mantere, Timo and Turunen, Pekka. :
Genetic algorithm based software testing.
Turku Centre for Computer Science, Springer-Verlag (1998)

\bibitem {7}
Alba,, Enrique and Chicano,, Francisco. :
Observations in using parallel and sequential 
evolutionary algorithms for automatic software testing.
Comput. Oper. Res., 3161--3183 (2008)

\bibitem {8}
Ribeiro,, Jos\'{e} Carlos Bregieiro :
Search-based test case generation for object-oriented java
 software using strongly-typed genetic programming.
GECCO '08: Proceedings of the 2008 GECCO conference companion on Genetic and evolutionary computation, 1819--1822 (2008)

\bibitem {9}
Harmen-Hinrich Sthamer :
The Automatic Generation of Software Test Data Using
 Genetic Algorithms.
University of Glamorgan (1995)

\bibitem {10}
Mark Last,, Shay Eyal,, Abraham Kandel :
Effective Black-Box Testing with Genetic Algorithms.
Hardware and Software, Verification and Testing, 134--148 (2006)

\bibitem {11}
Bechhofer,, Robert E.:
Selection in factorial experiments.
WSC '77: Proceedings of the 9th conference on Winter simulation 65--70 (1977)

\bibitem {12}
O'Neill,, Michael and Hemberg,, Erik and Gilligan,, Conor and Bartley,, Eliott and McDermott,, James and Brabazon,, Anthony:
GEVA: grammatical evolution in Java.
SIGEVOlution,ACM, 17--22 (2008)

\bibitem {13}
Klaus Meffert , Neil Rotstan:
JGAP: Java Genetic Algorithms Package.

\bibitem {14}
Wu,, Lieh-Ming and Wang,, Kuochen and Chiu,, Chuang-Yi:
A BNF-based automatic test program generator 
for compatible microprocessor verification
ACM Trans. Des. Autom. Electron. Syst., 105--132 (2004)

\bibitem {15}
BMiyake, J. and Brown, G.:
Automatic test generation for functional 
verification of microprocessors.
In Proceedings of the Third Asian Test Symposium , 292--297 (1994)

\bibitem {16}
Ciupa,, Ilinca and Leitner,, Andreas and Oriol,, Manuel and Meyer,, Bertrand.:
Experimental assessment of random testing
 for object-oriented software.
ISSTA '07: Proceedings of the 2007 international symposium on Software testing and analysis, 84--94 (2007)

\bibitem {17}
Gary Mcgraw and Christoph Michael and Michael Schatz:
Generating Software Test Data by Evolution.
IEEE Transactions on Software Engineering, 1085--1110 (1997)

\bibitem {18}
Yuesheng Qi,, Baozhong Wang,, Lishan Kang. :
Genetic Programming with simple loops.
Journal of Computer Science and Technology,429--434 (1999)

\bibitem {19}
Wolfgang Stöcher:
Designing and Prototyping a Functor
 Language Using Denotational Semantics.
RISC Report Series, University of Linz, Austria, (1997)

\bibitem {20}
Derk, M. D.:
Towards a simpler method of 
operational semantics for language definition.
SIGPLAN Not., ACM, 39--44 (2005)

\bibitem {21}
Bottaci, Leonardo.:
Genetic Algorithm Fitness 
Function for Mutation Testing.
Department of Computer Science, The University of Hull, Hull HU6 7RX, U.K. (2001)


\end{thebibliography}
\end{document}